\documentclass[preprint,showpacs,preprintnumbers,amsmath,amssymb,amssymb]{revtex4}

\usepackage{graphicx}% Include figure files
\usepackage{dcolumn}% Align table columns on decimal point
\usepackage{bm}% bold math
\usepackage{epstopdf}
\DeclareMathOperator{\BR}{BR}
\begin{document}

\hyphenpenalty=5000
\tolerance=1000
\makeatletter

%\preprint{APS/123-QED}

\title{Neutral 3-3-1 Higgs Boson Through $e^{+}e{-}$ Collisions }

\author{J.\ E.\ Cieza Montalvo$^1$}
\affiliation{$^1$Instituto de F\'{\i}sica, Universidade do Estado do Rio de Janeiro, Rua S\~ao Francisco Xavier 524, 20559-900 Rio de Janeiro, RJ, Brazil}
\author{C. A. Morgan Cruz, R. J. Gil Ram\'{i}rez, G. H. Ram\'{i}rez Ulloa, A. I. Rivasplata Mendoza$^2$}
\affiliation{$^2$Universidad Nacional de Trujillo, Departamento de F\'{i}sica, Av. Juan Pablo II S/N; Ciudad Universitaria, Trujillo, La Libertad, Per\'u}
\author{M. D. Tonasse$^{3}$\footnote{Permanent address: Universidade Estadual Paulista, {\it Campus} Experimental de Registro, Rua Nelson Brihi Badur 430, 11900-000 Registro, SP, Brazil}}
\affiliation{$^3$Instituto de F\'\i sica Te\'orica, Universidade Estadual Paulista, \\  Rua Dr. Bento Teobaldo Ferraz 271, 01140-070 S\~ao Paulo, SP, Brazil}
\date{\today}

%Lines break
%automatically or can be forced with \\
% It is always\today, today,
% but any date may be explicitly specified
\begin{abstract}
In this work we present an analysis of production and signature of neutral Higgs bosons $H_2^0$ in the version of the 3-3-1 model containing heavy leptons at the ILC (International Linear Collider) and CLIC (Cern Linear Collider). The production rate is found to be significant for the direct production of $e^{-} e^{+} \rightarrow H_{2}^{0} Z$. We also studied the possibility to identify it using their respective branching ratios.

\end{abstract}

\maketitle

%%%%%%%%%%%%%%%%%%%%%%%%%%%%%%%%%%%%%%%%%%%%%%%%%%%%%%%%%%%%%%%%%%%%%%

%\section{INTRODUCTION}
\section{INTRODUCTION \label{introd}}

The Higgs sector still remains one of the most indefinite part of the standard model (SM) \cite{wsg}, but it still represents a fundamental rule by explaining how the particles gain masses by means of a isodoublet scalar field, which is responsible for the spontaneous breakdown of the gauge symmetry, the process by which the spectrum of all particles are generated. This process of mass generation is the so called {\it Higgs mechanism}, which plays a central role in gauge theories. \par

The SM provides a very good description of all the phenomena related to hadron and lepton colliders. This includes the Higgs boson which appears as elementary scalar and which arises through the breaking of electroweak symmetry. The Higgs Boson is an important prediction of several quantum field theories and is so crucial to our understanding of the Universe. So on 4 July 2012, was measured the discovered 126 GeV Higgs boson \cite{atlas1, atlas11}. In this model, the Higgs field receives a vacuum expectation value (VEV), $v \simeq 246$ GeV, which breaks the electroweak gauge symmetry and gives masses to the fundamental fermions and gauge bosons. \par

However, the standard model does not predict the number of scalar multiplets of the theory, for that reason, there are several extensions of the standard model containing neutral and charged Higgs bosons. Since the standard model leaves many questions open, there are several well motivated extensions  of it. For example, if the Grand Unified Theory (GUT) contains the standard model at high energies, then the Higgs bosons associated with GUT symmetry breaking must have masses of order $M_{X} \sim {\cal O} (10^{15})$ GeV. Supersymmetry \cite{supers} provides a solution to this hierarchy problem through the cancellation of the quadratic divergences via the contributions of fermionic and bosonic loops \cite{cancell}. Moreover, the Minimal Supersymmetric extension of the Standard Model (MSSM) can be derived as an effective theory from supersymmetric Grand Unified Theories \cite{sgut}. Another promissory class of models is the one based on the $SU(3)_{C}\otimes  SU(3)_{L} \otimes U(1)_{N}$ (3-3-1 for short) semisimple symmetry group \cite{PT93}. In this model the new leptons do not require new generations, as occur in most of the heavy-lepton models \cite{FH99}. This ones is a chiral electroweak model whose left-handed charged heavy-leptons, which we denote by $P_a$ $=$ $E$, $M$ and $T$, together with the associated ordinary charged leptons and its respective neutrinos, are accommodated in SU(3)$_L$ triplets. \par

These models emerge as an alternative solution to the problem of violation of unitarity at high energies in processes such as $e^-e^- \to W^-V^-$, induced by right-handed currents coupled to a vector boson $V^-$. The usual way to circumvent this problem is to give particular values to model parameters in order to cancel the amplitude of the process \cite{PP92}, but in this work was proposed an elegant solution assuming the presence of a doubly charged vector boson. The simplest electroweak gauge model is able to realize naturally a double charge gauge boson based on the SU(3)$\otimes$U(1) symmetry \cite{PP92}. As a consequence of the extended gauge symmetry, the model is compelled to accommodate a much richer Higgs sector.  \par

The main feature of the 3-3-1 model is that it is able to predicts the correct number of fermions families. This is because, contrary to the standard model, the 3-3-1 model is anomalous in each generation. The anomalies are cancelled only if the number of families is a multiple of three. In addition, if we take into account that the asymptotic freedom condition of the QCD is valid only if the number of generations of quarks is to be less than five, we conclude that the number of generations is three \cite{LS01}.  Another good feature is that the model predicts an upper bound for the Weinberg mixing angle at $\sin^{2} {\theta_W} < 1/4$. Therefore, the evolution of $\theta_W$ to high values leads to an upper bound to the new mass scale between 3 TeV and 4 TeV \cite{JJ97}. \par
	
In this work we are interested in a version of the 3-3-1 model, whose scalar sector has only three Higgs triplets \cite{PT93}. The text is organized as follow. In Sect.\ref{sec2} we give the relevant features of the model. In Sect.\ref{sec3} we compute the total cross sections of the process $e^{-} e^{+} \rightarrow H_{2}^{0} Z$ and the Sect.\ref{sec4} contains our results and conclusions. \par

%%%%%%%%%%%%%%%%%%%%%%%%%%%%%%%%%%%%%%%%%%%%%%%%%%%%%%%%%%%%%%%%%%%%%

\section{Basic facts about the 3-3-1 model } \label{sec2}

The three Higgs triplets of the model are
%\begin{subequations}
\begin{eqnarray}
\eta & = & \left(\begin{array}{c} \eta^0 \\  \eta_1^- \\  \eta_2^+ \end{array}\right) \quad \rho = \left(\begin{array}{c} \rho^+ \\  \rho^0 \\  \rho^{++}
\end{array}\right)  \quad  \chi  =  \left(\begin{array}{c} \chi^-   \\
\chi^{--} \\ \chi^0 \end{array}\right)
%\label{higgs}
\end{eqnarray}
%\end{subequations}
transforming as $\left({\bf 3}, 0\right)$, $\left({\bf 3}, 1\right)$ and $\left({\bf 3}, -1\right)$, respectively.

The neutral scalar fields develop the vacuum expectation values (VEVs) $\langle\eta^0\rangle \equiv v_\eta$, $\langle\rho^0\rangle \equiv v_\rho$ and  $\langle\chi^0\rangle \equiv v_\chi$, with $v_\eta^2 + v_\rho^2 = v_W^2 = (246 \mbox{ GeV})^2$. The pattern of symmetry breaking is
$\mbox{SU(3)}_L \otimes\mbox{U(1)}_N \stackrel{\langle\chi\rangle}{\longmapsto}\mbox{SU(2)}_L\otimes\mbox{U(1)}_Y\stackrel{\langle\eta, \rho\rangle}{\longmapsto}\mbox{U(1)}_{\rm em}$
and so, we can expect $v_\chi \gg v_\eta, v_\rho$. The $\eta$ and $\rho$ scalar triplets give masses to the ordinary fermions and gauge bosons, while the $\chi$ scalar triplet gives masses to the new fermions and new gauge bosons. The most general, gauge invariant and renormalizable Higgs potential is
%\begin{widetext}

\begin{eqnarray}
V\left(\eta, \rho, \chi\right) & = & \mu_1^2\eta^\dagger\eta + \mu_2^2\rho^\dagger\rho + \mu_3^2\chi^\dagger\chi + \lambda_1\left(\eta^\dagger\eta\right)^2 + \lambda_2\left(\rho^\dagger\rho\right)^2 +       \lambda_3\left(\chi^\dagger\chi\right)^2 +   \nonumber     \\
&& \left(\eta^\dagger\eta\right)\left[\lambda_4\left(\rho^\dagger\rho\right) + \lambda_5\left(\chi^\dagger\chi\right)\right] +
+ \lambda_6\left(\rho^\dagger\rho\right)\left(\chi^\dagger\chi\right) + \lambda_7\left(\rho^\dagger\eta\right)\left(\eta^\dagger\rho\right) +   \nonumber  \\
&& \lambda_8\left(\chi^\dagger\eta\right)\left(\eta^\dagger\chi\right) +   \lambda_9\left(\rho^\dagger\chi\right)\left(\chi^\dagger\rho\right) + \lambda_{10}\left(\eta^\dagger\rho\right)\left(\eta^\dagger\chi\right) +   \nonumber    \\
&& \frac{1}{2}\left(f\epsilon^{ijk}\eta_i\rho_j\chi_k + {\mbox{H. c.}}\right).
\label{pot}\end{eqnarray}

%\end{widetext}
Here $\mu_i$ $\left(i = 1, 2, 3\right)$, $f$ are constants with dimension of mass and the $\lambda_i$, $\left(i = 1, \dots, 10\right)$ are dimensionless constants. $f$ and $\lambda_3$ are negative from the positivity of the scalar masses. The term proportional to $\lambda_{10}$ violates lepto-barionic number,
therefore it was not considered in the analysis of the Ref. \cite{TO96} (another analysis of the 3-3-1 scalar sector are given in Ref. \cite{AK} and references cited therein). We can notice that this term contributes to the mass matrices of the charged scalar fields, but not to the neutral ones.  However, it can be checked that in the approximation $v_\chi \gg v_\eta, v_\rho$ we can still work with the masses and eigenstates given in Ref. \cite{TO96}. Here this term is important to the decay of the lightest exotic fermion. Therefore, we will keep it in the Higgs potential (\ref{pot}).

As usual, symmetry breaking is implemented by shifting the scalar neutral fields $\varphi = v_\varphi + \xi_\varphi + i\zeta_\varphi$, with $\varphi$ $=$  $\eta^0$, $\rho^0$, $\chi^0$. Thus, the physical neutral scalar eigenstates  $H^0_1$, $H^0_2$, $H^0_3$ and $h^0$ are related to the shifted fields as

%\begin{subequations}
\begin{eqnarray}
\left(\begin{array}{c} \xi_\eta \\  \xi_\rho \end{array}\right) \approx
\frac{1}{v_W}\left(\begin{array}{cc} v_\eta & v_\rho \\  v_\rho & -v_\eta
\end{array}\right)\left(\begin{array}{c} H^0_1 \\  H^0_2 \end{array}\right),&& \\
\xi_\chi \approx H^0_3, \qquad \zeta_\chi  \approx h^0,&&
\label{eign}\end{eqnarray}
and in the charge scalar sector we have
\begin{eqnarray}
\eta^+_1 \approx \frac{v_\rho}{v_W}H^+_1, \qquad \rho^+ \approx \frac{v_\eta}{v_W}H_2^+, && \\ \chi^{++} \approx \frac{v_\rho}{v_\chi}H^{++}, &&
\label{eigc}\end{eqnarray}\label{eig}
%\end{subequations}
with the condition that $v_\chi \gg v_\eta, v_\rho$ \cite{TO96}. \par
The content of matter fields form the three SU(3)$_L$ triplets

%\begin{subequations}
\begin{eqnarray}
\psi_{aL} = \left(\begin{array}{c} \nu^\prime_{\ell a} \\ \ell^\prime_a \\ P^\prime_a  \end{array}\right), \nonumber  &&  \\ Q_{1L} = \left(\begin{array}{c} u^\prime_1 \\ d^\prime_1 \\ J_1  \end{array}\right), \qquad Q_{\alpha L} = \left(\begin{array}{c} J^\prime_\alpha \\ u^\prime_\alpha \\ d^\prime_\alpha  \end{array}\right), &&
\label{fer}\end{eqnarray}
%\end{subequations}
transform as $\left({\bf 3}, 0\right)$, $\left({\bf 3}, 2/3\right)$ and $\left({\bf 3}^*, -1/3\right)$, respectively, where $\alpha = 2, 3$. In Eqs. (\ref{fer}) $P_a$ are heavy leptons, $\ell^\prime_a = e^\prime, \mu^\prime, \tau^\prime$. The model also predicts the exotic $J_1$ quark, which carries $5/3$ units of elementary electric charge and $J_2$ and $J_3$ with $-4/3$ each. The numbers $0$, $2/3$ and $-1/3$ in Eqs. (\ref{fer}) are the U$_N$ charges. We also have
the right-handed counterpart of the left-handed matter fields, $\ell^\prime_R \sim \left({\bf 1}, -1\right)$, $P^\prime_R \sim \left({\bf 1}, 1\right)$, $U^\prime_R \sim \left({\bf 1}, 2/3\right)$, $D^\prime_R \sim \left({\bf 1}, -1/3\right)$, $J^\prime_{1R} \sim \left({\bf 1}, 5/3\right)$ and $J^\prime_{2,3R} \sim \left({\bf 1}, -4/3\right)$, where $U = u, c, t$ and $D = d, s, b$ for the ordinary quarks. \par
The Yukawa Lagrangians that respect the gauge symmetry are
%\begin{widetext}
%\begin{subequations}
\begin{eqnarray}
{\cal L}^Y_\ell & = & -G_{ab}\overline{\psi_{aL}}\ell^\prime_{bR} - G^\prime_{ab}\overline{\psi^\prime_{aL}}P^\prime\chi + {\mbox{H. c.}}, \\
{\cal L}^Y_q & = & \sum_a\left[\overline{Q_1{L}}\left(G_{1a}U^\prime_{aR}\eta + \tilde{G}_{1a}D^\prime_{aR}\rho\right) + \sum_\alpha\overline{Q_{\alpha L}}\left(F_{\alpha a}U^\prime_{aR}\rho^* + \tilde{F}_{\alpha a}D^\prime_{aR}\eta^*\right)\right] +  \cr &&  +\sum_{\alpha\beta}F^J_{\alpha\beta}\overline{Q_{\alpha J}}J^\prime_{\beta R}\chi^* + G^J\overline{Q_{1L}}J_{1R} + {\mbox{ H. c.}}.
\label{yuk}
\end{eqnarray}

%\end{subequations}\end{widetext}
Here, the $G$'s, $\tilde{G}$'s, $F$'s and $\tilde{F}$'s are Yukawa coupling constants with $a, b = 1, 2, 3$ and $\alpha = 2, 3$. \par
It should be noticed that the ordinary quarks couple only through $H^0_1$ and $H^0_2$. This is because these physical scalar states are linear combinations of the interactions eigenstates $\eta$ and $\rho$, which break the SU(2)$_L$  $\otimes$U(1)$_Y$ symmetry  to U(1)$_{\rm em}$. On the other hand the heavy-leptons and quarks couple only through $H^0_3$ and $h^0$ in scalar sector, {\it i. e.}, throught the Higgs that induces the symmetry breaking of SU(3)­$_L$$\otimes$U(1)$_N$ to SU(2)$_L$$\otimes$U(1)$_Y$. The Higgs particle spectrum consists of ten physical states: three scalars ($H_{1}^{0}, H_{2}^{0}, H_3^{0}$), one neutral pseudoscalar $h^0$ and six charged Higgs bosons, $H_{1}^{\pm}, H_{2}^{\pm}, H^{\pm\pm}$. \par

In this work we study the production of a neutral Higgs boson $H_2^0$, which can be radiated from a $Z^{'}$ boson at $e^{+} e^{-}$ colliders such as the International Linear Collider  (ILC) ($\sqrt{s} = 1500$ GeV) and CERN Linear Collider (CLIC) ($\sqrt{s} = 3000$ GeV).

%%%%%%%%%%%%%%%%%%%%%%%%%%%%%%%%%%%%%%%

%%%%%%%%%%%%%%%%%%%%%%%%%%%%%%%%%%%%%%%%%%%%%%%%%%%%%%%%%%%%%%%%%%%%%

\section{CROSS SECTION PRODUCTION}\label{sec3}

We begin with the direct production of Higgs ($H_{2}^{0}$), that is $e^{-} e^{+} \rightarrow H_{2}^{0} Z$. This process take place via the exchange of a virtual $Z^{\prime}$ boson in the s channel and it can also take place through the $H_{1}^{0}$ and $H_{2}^{0}$, but the contribution of these channels are small due to the small coupling of the Higgs $H_{2}^{0}$ to the electrons. The term involving the $Z$ boson is absent, because there is no coupling between the $Z$ and $H_{2}^{0} Z$. Then using the interaction Lagrangian Eqs. ($2$) and ($10$) we obtain the differential cross section.

\begin{eqnarray}
\left (\frac{d \hat{\sigma}}{d\cos \theta} \right )_{H_{2}^{0} Z} & = &\frac{\beta_{H_{2}^{0}} \alpha^{2} \pi}{32 \sin^{4}_{\theta_{W}} \cos^{2}_{\theta_{W}} s} \ \frac{\Lambda_{ZZ^{\prime} H_{2}^{0}}^2}{(s- M_{Z'}^{2}+ iM_{Z'} \Gamma_{Z'})^{2}}
\Biggl  \{  (2M_{Z}^{2}+ \frac{2tu}{M_{Z}^{2}}- 2t- 2u + 2s)  \nonumber  \\
&& (g_{V'}^{e^{2}}+ g_{A'}^{e^{2}}) \Biggr \} , \nonumber  \\
\label{DZZ'H}
\end{eqnarray}
%\end{widetext}
the $\beta_{H_{2}^{0}}$ is the Higgs velocity in the c.m. of the subprocess which is equal to
\[
\beta_{H_{2}^{0}} = \frac{ \left [\left( 1- \frac{(m_{Z}+ m_{H_{2}^{0}})^{2}}{\hat{s}} \right) \left(1- \frac{(m_{Z}- m_{H_{2}^{0}})^{2}}{\hat{s}} \right) \right ]^{1/2}}{1-\frac{m_{Z}^{2}-m_{H_{2}^{0}}^{2}}{\hat{s}}}  \ \ ,
\]

and $t$ and $u$ are

\[
t  = m_{Z}^{2} - \frac{s}{2} \Biggl \{ \left(1+ \frac{m_{Z}^{2}- m_{H}^{2}}{s}\right)- \cos \theta  \left [\left( 1- \frac{(m_{Z}+ m_{H})^{2}}{s} \right) \left(1- \frac{(m_{Z}- m_{H})^{2}}{s} \right) \right ]^{1/2}\Biggr \},
\]

\[
u  = m_{H}^{2} - \frac{s}{2} \Biggl \{ \left(1- \frac{m_{Z}^{2}- m_{H}^{2}}{s}\right)+ \cos \theta  \left [\left( 1- \frac{(m_{Z}+ m_{H})^{2}}{s} \right) \left(1- \frac{(m_{Z}- m_{H})^{2}}{s} \right) \right ]^{1/2}\Biggr \},
\]
where $\theta$ is the angle between the Higgs and the incident quark in the CM frame. \par

The primes $\left(^\prime\right)$ are for the case when we take a $Z'$ boson, $\Gamma_{Z'}$ \cite{ct2005, cieto02}, are the total width of the $Z'$ boson, $g_{V', A'}^{e}$ are the 3-3-1 lepton coupling constants, $s$ is the center of mass energy of the $e^{-} e^{+}$ system, $g= \sqrt{4 \ \pi \ \alpha}/\sin \theta_{W}$ and $\alpha$ is the fine structure constant, which we take equal to $\alpha=1/128$. For the $Z^\prime$ boson we take  $M_{Z^\prime} = \left(1.5 - 3\right)$ TeV, since $M_{Z^\prime}$ is proportional to the VEV $v_\chi$ \cite{TO96,PP92,fra92}. For the standard model parameters, we assume Particle Data Group values, {\it i. e.}, $M_Z = 91.19$ GeV, $\sin^2{\theta_W} = 0.2315$, and $M_W = 80.33$ GeV  \cite{Nea10}, $\it{t}$ and $\it{u}$ are the kinematic invariants. We have also defined the $\Lambda_{ZZ^{\prime} H_{2}^{0}}$ as the coupling constants of the $Z^{\prime}$ boson to Z boson and Higgs $H_{2}^{0}$, and the $\Lambda_{e \bar{e} Z^{\prime}}$ are the coupling constants of the $Z^{\prime}$ to $e \bar{e}$. \par

\begin{subequations}\begin{eqnarray}
\left(\Lambda_{e\bar{e}Z^\prime}\right)_\mu & \approx & -i\frac{g}{2 \sqrt{1-s_w^2}} \gamma_\mu\left[g_{V^\prime}^{e}  - g_{A^\prime}^{e} \gamma_5\right], \\
\left(\Lambda_{ZZ^\prime H_2^0}\right)_{\mu\nu} & \approx & \frac{g^2}{\sqrt{3}\left(1 - 4s_W^2\right)}\frac{v_\eta v_\rho}{v_W}g_{\mu\nu},
\label{eigc}\end{eqnarray}\label{eigthen}\end{subequations}

%%%%%%%%%%%%%%%%%%%%%%%%%%%%%%%%%%%%%%%%%%%%%%%%%%%%%%%%%%%%%%%%%
%%%%%%%%%%%%%%%%%%%%%%%%%%%%%%%%

\section{RESULTS AND CONCLUSIONS}\label{sec4}

Here we present the cross section for the process $e^+ e^-  \rightarrow H_2^0 Z$ for the ILC ($1.5$) TeV and CLIC ($3$ TeV). All calculations were done according to \cite{TO96,cnt2} from which we obtain for the parameters and the VEV, the following representative values:  $\lambda_{1} =0.3078$,  $\lambda_{2}=1.0$, $\lambda_{3}= -0.025$, $\lambda_{4}= 1.388$, $\lambda_{5}=-1.567$, $\lambda_{6}= 1.0$, $\lambda_{7} =-2.0$, $\lambda_{8}=-0.45$,  $v_{\eta}=195$ GeV, and $\lambda_{9}=-0.90(-0.76,-0.71)$ correspond to $v_\chi= 1000(1500,2000)$ GeV these parameters and VEV are used to estimate the values for the particle masses which are given in table \ref{tab1}.  \par

Differently from what we did in the paper \cite{ct2005}, where was taken arbitrary parameters, in this work we take for the parameters and the VEV the following representative values given above and also the fact that the mass of $m_{H_1^0}$ is already defined \cite{atlas1, atlas11}. It is remarkable that the cross sections were calculated in order to guarantee the approximation $-f \simeq v_\chi$ \cite{TO96, cnt2}.  It must be taken into consideration that the branching ratios of $H_2^0$ are dependent on the parameters of the 3-3-1, which determines the size of several decay modes. \par
	
\begin{table}[h]
\caption{\label{tab1} Values for the particle masses used in this work.
All the values in this Table are given in GeV. Here, $m_{H^{\pm\pm}} =
500$ GeV and $m_T = 2v_\chi$.}
\begin{ruledtabular}
\begin{tabular}{c|ccccccccccccccc}
$f$ & $v_{\chi}$, $m_{J_1}$ & $m_E$ & $m_M$  & $m_{H_3^0}$ & $m_{h^0}$ &
$m_{H_1^0}$ & $m_{H_2^0}$ & $m_{H^\pm_2}$ & $m_V$ & $m_U$ & $m_{Z^\prime}$
& $m_{J_{2, 3}}$ \\
\hline
-1008.3 & 1000 & 148.9 & 875       & 2000   & 1454.6   & 126     & 1017.2
 & 183      & 467.5    & 464 & 1707.6 &1410 \\
-1499.7 & 1500 & 223.3 & 1312.5  & 474.34 & 2164.32 & 125.12 & 1525.8   &
387.23 &  694.12 & 691.76 & 2561.3 & 2115 \\
-1993.0 & 2000 & 297.8 & 1750     & 632.45 & 2877.07 & 125.12 & 2034.37 &
519.39 & 922.12  & 920.35 & 3415.12 & 2820 \\
\end{tabular}
\end{ruledtabular}
\end{table}

The Higgs $H_{2}^{0}$ in 3-3-1 model is not coupled to a pair of standard bosons, it couples to quarks, leptons, Z $Z^{\prime}$, $Z^{\prime}$ $Z^{\prime}$
gauge bosons,  $H_{1}^{-} H_{1}^{+}$, $H_{2}^{-} H_{2}^{+}$, $h^{0}   h^{0}$, $H_{1}^{0}  H_{3}^{0}$ Higgs bosons, $V^{-}V^{+}$ charged bosons, $U^{--} U^{++}$ double charged bosons, $H_{1}^{0} Z$, $H_{1}^{0} Z'$ bosons and $H^{--} H^{++}$ double charged Higgs bosons \cite{ct2005}. The Higgs $H_{2}^{0}$ can be much heavier than $ 1017.2$ GeV for $v_\chi = 1000 \ $GeV, $1525.8$ GeV for $v_\chi = 1500 \ $GeV, and $2034.37$ GeV for $v_\chi = 2000$ \ GeV, so the Higgs $H_2^{0}$ is a heavy particle. \par

In Table \ref{tab1} the masses of the exotic boson $Z^{\prime}$, taken above, is in accord with the estimates of the Tevatron, which probes the $Z^{\prime}$ masses in the 923-1023 GeV range, \cite{tait}, while the reach of the LHC is superior for higher masses, that is $1 \ TeV <M_{Z^{\prime}} \leq 5$ TeV \cite{freitas} and for ATLAS at $8$ TeV with an integrated luminosity approximately of $20$ fb$^{-1}$ the mass range is $2 \ TeV \leq m_{Z^{\prime}} \leq 3$ TeV, \cite{atlas1,atlas2014}.

\subsection{ILC - Events}

\begin{figure}
\begin{center}
\includegraphics{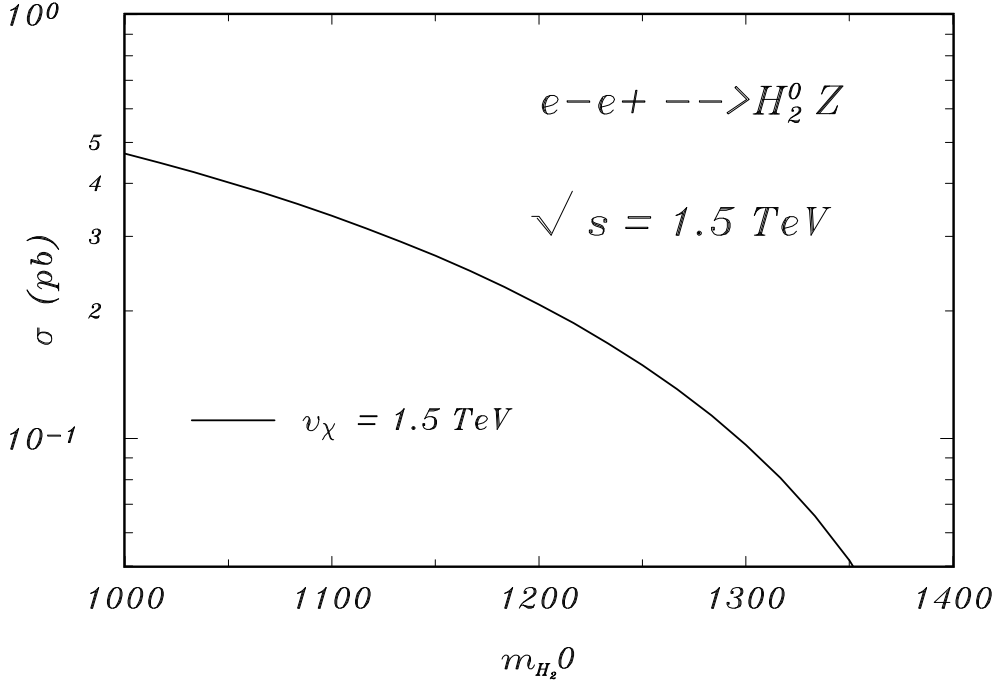}
\end{center}
\caption{Total cross section for the process $e^+ e^-  \rightarrow H_2^0 Z$ as a function of $m_{H^{0}_{2}}$ for the ILC at 1.5 TeV and $v_{\chi}=1.5$ TeV. }
\label{fig1}
\end{figure}

Considering that the expected integrated luminosity for ILC collider will be of order of $500$ fb$^{-1}$, then the statistics we are expecting are the following, the ILC gives a total of $ \simeq 1.68 \times 10^5 (4.83 \times 10^4)$ events per year, if we take the mass of the Higgs boson $m_{H_2^0}= 1100(1300)$ GeV ($\Gamma_{H_2^0} = 878.25, 1091.33 GeV$) and $v_{\chi}=1000$ GeV, see Fig. \ref{fig1}. These values are in accord with the Table \ref{tab1}.

To obtain event rates we multiply the production cross sections by the respective branching ratios. Considering that the signal for  $H_{2}^{0}Z$ production for  $m_{H_{2}^{0}}= 1100(1300)$ GeV and $v_{\chi}=1000$ GeV will be $H_{2}^{0} Z \rightarrow  Z H_{1}^{0} Z$, and taking into account that the branching ratios for these particles would be $BR(H^{0}_{2} \to Z H_{1}^{0}) = 39.5 (43.4)  \ \% $ \cite{ctrg2013}, and $\BR(Z \to  b \bar{b}) = 15.2 \ \% $, and that the particles $H_{1}^{0}$ decay into  $W^{+} W^{-}$, and taking into account that the branching ratios for these  particles would be $BR(H^{0}_{1} \to W^{+} W^{-}) = 23.1 \ \% $ followed by leptonic decay of the boson $W^{+}$ into $\ell^{+} \nu$ and $W^{-}$ into $\ell^{-} \bar{\nu}$ whose branching ratios for these particles would be  $BR(W  \to \ell \nu) = 10.8 \ \%$, then we would have approximately  $ \simeq 4 (1))$  events per year for ILC for the signal $b\bar{b} b \bar{b} \ell^{+} \ell^{-} X$. \par

Statistics for $v_{\chi}=1500 (2000)$ gives no result because there is not enough energy to produce the Higgs boson $H_2^0$. That is, we can see that the number of events for the signal for ILC is insignificant.  \par

\subsection{CLIC - Events}

\begin{figure}
\begin{center}
\includegraphics[width=1.1\columnwidth]{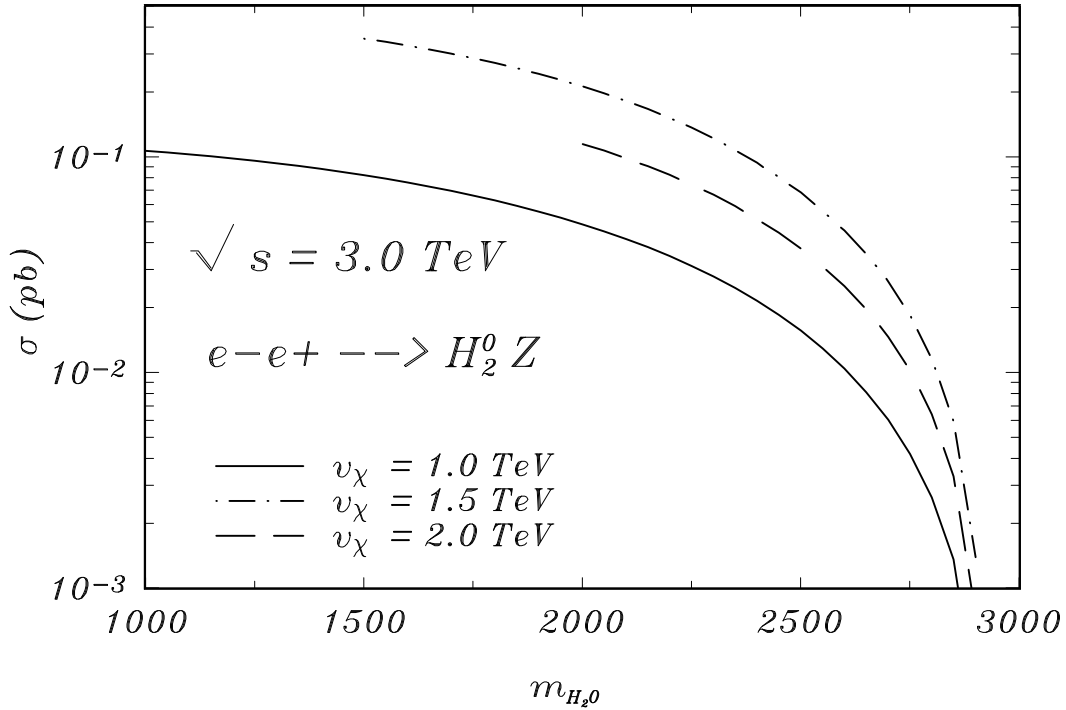}
\end{center}
\caption{Total cross section for the process $e^+ e^-  \rightarrow H_2^0 Z$ as a function of $m_{H^{0}_{2}}$ for the CLIC at 3.0 TeV and $v_{\chi}=1.0$ TeV (solid line), $v_\chi = 1.5$ TeV (dash-dot line), $v_\chi = 2.0$ TeV (dashed line).}
\label{fig2}
\end{figure}

Considering that the expected integrated luminosity for CLIC collider will be of order of $3000$ fb$^{-1}$/yr, then we obtain a total of $ \simeq 3.1 \times 10^5 (2.8) \times 10^5$ events per year if we take the mass of the Higgs boson $m_{H_2^0}= 1100(1300)$ GeV and  $v_{\chi}=1000$ GeV, see Fig.\ref{fig2}. Considering the same signal as above for $H_2^0 Z$ production, that is $H_{2}^{0} Z \rightarrow  Z H_{1}^{0} Z$, and taking into account that the branching ratios for these particles would be $BR(H^{0}_{2} \to Z H_{1}^{0}) = 39.5 (43.4)  \ \% $, \cite{ctrg2013}, and $\BR(Z \to  b \bar{b}) = 15.2 \ \% $, and that the particles $H_{1}^{0}$ decay into  $W^{+} W^{-}$, and taking into account that the branching ratios for these  particles would be $\BR(H^{0}_{1} \to W^{+} W^{-}) = 23.1 \ \% $ followed by leptonic decay of the boson $W^{+}$ into $e^{+} \nu$ and $W^{-}$ into $e^{-} \bar{\nu}$ whose branching ratios for these particles would be  $BR(W  \to e \nu) = 10.8 \ \%$, then we would have approximately  $ \simeq 8(8)$  events per year for CLIC for the signal $b\bar{b} b \bar{b} \ell^{+} \ell^{-} X$. \par
The statistics for $v_{\chi}=1500$ gives a total of  $\simeq 9.8 \times 10^5(8.1 \times 10^5)$ events per year for CLIC, if we take the mass of the Higgs boson $m_{H_{2}^{0}}= 1600(1800)$ GeV, respectively. These values are in accord with Table \ref{tab1}. Taking into account the same signal as above, that is $H_{2}^{0} Z \rightarrow  Z H_{1}^{0} Z$, and taking into account that the branching ratios for these particles would be $\BR(H^{0}_{2} \to Z H_{1}^{0}) = 44.2 (45.9)  \ \% $, \cite{ctrg2013}, $BR(Z \to  b \bar{b}) = 15.2 \ \% $, $BR(H^{0}_{1} \to W^{+} W^{-}) = 23.1 \ \% $, $BR(W  \to e \nu) = 10.8 \ \%$, we would have approximately  $ \simeq 27(23))$  events per year for CLIC for the same signal $b\bar{b} b \bar{b} e^{+} e^{-} X$. \par

With respect to vacuum  expectation  value $v_{\chi}=2000$ GeV, for the masses of $m_{H_{2}^{0}}= 2100(2300)$ it  will give a total of  $\simeq 2.9 \times 10^5(2.0 \times 10^5 )$  events per year to produce $H_{2}^{0}$.  Taking into account the same signal as above, that is $b\bar{b} b \bar{b} \ell^{+} \ell^{-} X$ and considering  that the branching ratios for $H_{2}^{0}$ would be $\BR(H^{0}_{2} \to Z H_{1}^{0}) = 46.4 (47.3)  \ \% $, \cite{ctrg2013},  $\BR(Z \to  b \bar{b}) = 15.2 \ \% $, $\BR(H^{0}_{1} \to W^{+} W^{-}) = 23.1 \ \% $, $BR(W  \to e \nu) = 10.8 \ \%$, we will have approximately $ \simeq 8(6)$ events per year \par

The main background to this signal is $Z W^{+} W^{-} Z$, which cross section is $1.17 \times 10^{-3}$ pb for $\sqrt{s}=3$ TeV. Considering that the $Z Z$ particles decay into $b \bar{b}$, whose branching ratios for these particles would be $\BR(Z \rightarrow b \bar{b}) = 15.2 \%$ followed by leptonic decay of the boson W, that is $\BR(W \rightarrow e \nu) = 10.8 \%$ then we would have approximately a total of $ \simeq 1$ event for the background and $\simeq 8(8)$ events for the signal for $m_{H_{2}^{0}}= 1100(1300)$ GeV and $v_{\chi}=1000$. \par

Therefore we have that the statistical significance is $\simeq 2.66(2.66) \sigma$ for $m_{H_{2}^{0}}= 1100(1300)$ and $v_{\chi}=1000$ GeV, that is a low probability to detect signals. On the other hand, for $v_{\chi}=1500$ GeV and $m_{H_{2}^{0}}= 1600(1800)$ GeV we have $\simeq 5.10(4.70) \sigma$ discovery in the $b\bar{b} b \bar{b} e^{+} e^{-} X$ final state , for $v_{\chi}=2000$ GeV and $m_{H_{2}^{0}}= 2100(2300)$ GeV which corresponds to $\simeq 2.66(2.27) \sigma$, we have that the signals are too small to be observed. \par

To extract the signal from the background we must select the $b \bar{b}$ channel using the techniques of b-flavour identification. Later, the Z that comes together with the $H_{2}^{0}$ and the other Z that comes from the decay of $H_{2}^{0}$ would appear as a peak in the invariant mass distribution of b-quark pairs. The charged lepton track from the $W$ decay and the cut on the missing transverse momentum ${p\!\!\slash}_{T} >$ 20 GeV allows for a very strong reduction of the backgrounds. \par

The $H_{2}^{0} Z$ will also decay into  $t \bar{t} \ \ell^{+} \ell^{-}$, and consider that the branching ratios for these particles would be  $\BR(H^{0}_{2} \to  t \bar{t}) = 5.1 (4.1)  \ \% $, , \cite{ctrg2013}, and $BR(Z \to \ell^{+} \ell^{-}) = 10.2 \ \% $ for the mass of the Higgs boson $m_{H_{2}^{0}}= 1100(1300)$ GeV and $v_{\chi}=1000$ GeV and that the particles $t \bar{t}$ decay into $ b \bar{b} W^{+} W^{-}$, whose branching ratios for these particles would be  $BR(t \to b W) = 99.8 \ \% $, followed by leptonic decay of the boson W, that is  $BR(W \to e \nu) = 10.75 \ \% $, then  we would have approximately $\simeq 19 (13)$ events per year for CLIC for the signal $b\bar{b} e^{-} e^{+} \ell^{+} \ell^{-} X$. Considering the vacuum  expectation value $v_{\chi}=1500$ GeV and the branching ratios $\BR(H_{2}^{0} \rightarrow t \bar{t}) = 2.8 (2.3) \ \% $, , \cite{ctrg2013}, and taking the same parameters and branching ratios for the same particles given above, then we would have for  $m_{H_{2}^{0}}= 1600(1800)$ a total of  $ \simeq 32(22)$ events per year for CLIC for the same signal. With respect to vacuum  expectation  value $v_{\chi}=2000$ GeV, for the masses of $m_{H_{2}^{0}}= 2100(2300)$ and taking into account the same signal as above, that is $b\bar{b} e^{-} e^{+} \ell^{+} \ell^{-} X$ and considering  that the branching ratios $\BR(H_{2}^{0} \rightarrow t \bar{t}) = 1.7 (1.5) \ \% $, \cite{ctrg2013},we will have approximately $ \simeq 6(4)$ events per year. \par

Taking again the irreducible background for the process $t \bar{t}Z\rightarrow b \bar{b} e^{+} e^{-} \ell^{+} \ell^{-} X$, and using CompHep \cite{pukhov} we have that a cross section of $1.67 \times 10^{-3}$ pb, which gives $ \simeq 6$ events. So we will have a total of $\simeq  19 (13)$ events per year for the signals for $m_{H_{2}^{0}}= 1100(1300)$ GeV and $v_{\chi}=1000$, which corresponds to have $\simeq 3.80(2.98) \sigma$, then we have an evidence for $\simeq 3.80 \sigma$ discovery in the $b\bar{b} e^{-} e^{+} \ell^{+} \ell^{-} X$ final state. On the other hand, for $v_{\chi}=1500$ we have $\simeq 32(22)$ events for $m_{H_{2}^{0}}= 1600(1800)$ GeV and which corresponds to $\simeq 5.19(4.16) \sigma$, then we have a discovery for $\simeq 5.19 \sigma$ in the $b \bar{b} e^{+} e^{-} \ell^{+} \ell^{-} X$ final state. For $v_{\chi}=2000$ we have $\simeq 6(4)$ events for $m_{H_{2}^{0}}= 2100(2300)$ GeV and which corresponds to $\simeq 1.73(1.27) \sigma$, that is a low probability to detect the signals. We impose the following cuts to improve the statistical significance of a signal, i. e. we isolate a hard lepton from the $W$ decay with $p_{T}^{\ell}>$ 20 GeV, put the cut on the missing transverse momentum ${p\!\!\slash}_{T} >$ 20 GeV and apply the Z window cut $|m_{\ell^{+} \ell^{-}} - m_{Z}| >$ 10 GeV, which removes events where the leptons come from Z decay \cite{aguila}. However, all this scenarios can only be cleared by a careful Monte Carlo work to determine the size of the signal and background. \par

We still mention that the initial state radiation (ISR) and beamstrahlung (BS) strongly affects the behaviour of the production cross section around the resonance peaks, modifying as the shape as the size \cite{nicro}, so Fig. $3$ shows the cross section with and without ISR + BS around the resonance point $m_{Z^\prime} = 2561.3$ GeV for CLIC. As can be seen the peak of the resonance shifts to the right and is lowered as a  result of the ISR + BS effects. \par

\begin{figure}
\begin{center}
\includegraphics[width=1.1\columnwidth]{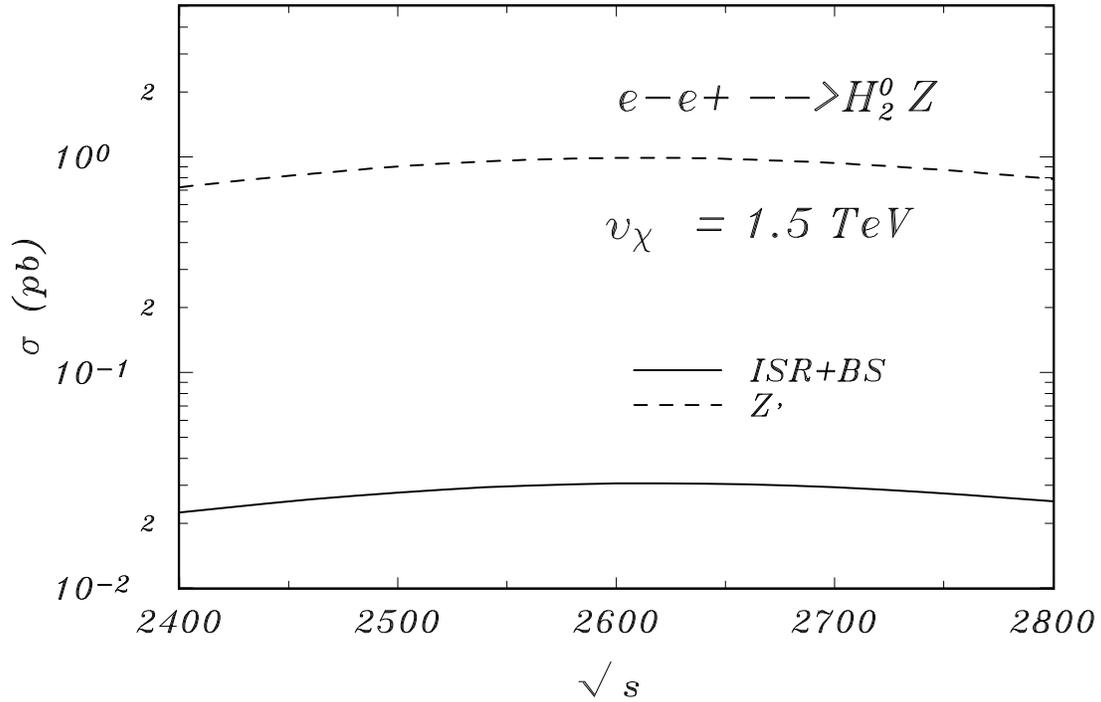}
\end{center}
\caption{Total cross section for the process $e^+ e^- \rightarrow H_2^0 Z$ as a function of center of mass energy ($\sqrt{s}$) with and without ISR + BS (dashed line and solid line respectively) for CLIC at $v_\chi=1.5$ TeV.}
\label{fig3}
\end{figure}

In summary, we showed in this work that in the context of the 3-3-1 model the signatures for neutral Higgs boson $H_2^0$ can be significant in CLIC collider if we take $v_{\chi}=1500$, $m_{H_{2}^{0}}=1600(1800)$ GeV and a luminosity of 3000 $fb^{-1}$, we have $\simeq 5.10(4.70) \sigma$ discovery in the $b\bar{b} b \bar{b} e^{+} e^{-} X$ and $\simeq 5.19(4.16) \sigma$ in the $b \bar{b} e^{+} e^{-} \ell^{+} \ell^{-} X$ final state. \par

\acknowledgments
{MDT is grateful to the Instituto de F\'\i sica Te\'orica of the UNESP for hospitality, the Brazilian agencies CNPq for a research grant, and FAPESP for financial support.}

\newpage

%^^^^^^^^^^^^^^^^^^^^^^^^^^^^^^^^^^^^^^^^^^^^^^^^^^^^^^^^^^^^^^^^^^^^^
%%  \section{Appendix}

%\begin{references}

\end{document}